\documentclass[12pt]{elsart}
\usepackage{graphicx}
\usepackage{epsf}

\begin{document}

\begin{frontmatter}

\title{Light transport in cold atoms: the fate of coherent backscattering in the weak localization regime}

\author[DW]{D. Wilkowski}
\author{Y. Bidel, T. Chaneli\`ere, R. Kaiser, B. Klappauf, G. Labeyrie,}
\author{C.A. M\"uller and Ch. Miniatura}

\address{Laboratoire Ondes et D\'esordre, FRE 2302 du CNRS, 1361 route des
Lucioles, Valbonne F-06560, France}

\thanks[DW]{\tt email wilkowsk@inln.cnrs.fr }


\begin{abstract}
The recent observation of coherent backscattering (CBS) of light by atoms
has emphasized the key role of the velocity spread and of the
quantum internal structure of the atoms. Firstly, using highly resonant
scatterers imposes very low temperatures of the disordered medium in order to
keep the full contrast of the CBS interference. This criterion is usually
achieved with standard laser cooling techniques. Secondly, a non trivial
internal
atomic structure leads to a dramatic decrease of the CBS contrast. Experiments
with Rubidium atoms (with a non trivial internal structure) and with Strontium
(with the simplest possible internal structure) show this behaviour and confirm
theoretical calculations.
\end{abstract}

\end{frontmatter}


After a few scattering mean free path $\ell$, a wave (with wavenumber $k=2\pi
/\lambda$) propagating in an opaque medium rapidly looses the memory of its
initial direction. At this scale, intensity propagation is often described as a
diffusion process. However this description discards an important phenomenon:
interference between multiply scattered waves. It is now known that
interference alters the wave transport and can, under suitable conditions,
bring it to a complete stop. This is the Anderson (or strong) localization
regime, where the diffusion is suppressed \cite{anderson}. For twenty years,
there have been tremendous experimental as well as theoretical efforts to study
interference effects in the multiple scattering regime \cite{multiple}.
\\ A hallmark in this field is the CBS cone which is observed as a
reflection peak with angular width $\simeq 1/k\ell$ at backscattering.
It corresponds to the incoherent
sum of two-waves interference between a possible {\it multiple} scattering
path amplitude
and its reversed counterpart \cite{barabanenkov}.
This interferential increase of the {\it configuration-averaged} diffuse
reflection off a disordered sample depends on the nature of the scatterer
but also, for vectorial waves like light, on the input/output polarization
channel. For
classical {\it spherically-symmetric} scatterers, the maximum enhancement
factor of
the CBS cone is 2 in the $h\|h$ helicity-preserving polarization channel,
independently of the geometrical shape of the scattering
medium \cite{wiersma95}. This is so for two reasons. First, in the
$h\|h$ channel single scattering is suppressed :  for spherically-symmetric
scatterers, the polarization is conserved at backscattering, like in specular
reflection off a mirror. Second, in the absence of a
magnetic field, the amplitudes of the
interfering paths are exactly equal at backscattering by virtue
of the reciprocity theorem \cite{Bart}.
\\ We will see that the observation of a CBS cone with a maximum contrast
on a resonant atomic gas,
however imposes severe restrictions on the external as well as
internal degrees of freedom of
the atoms. In section \ref{vi} we review some effects related to
{\it moving} resonant scatterers. In section \ref{internal} we show how
the internal structure of the atomic scatterer reduces the CBS contrast.
In section
\ref{res} experimental results, obtained in the weak localization regime
$k\ell\gg 1$, are presented. They confirm theoretical predictions about
the role of
the internal structure.

\section{CBS with resonant scatterers}\label{vi}

\subsection{Dynamical breakdown of CBS enhancement factor: Double scattering model}\label{vitesse}

\begin{figure}[h]
\begin{center}
\includegraphics[scale=0.5]{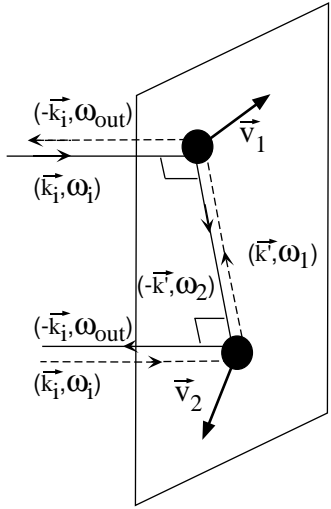}
\caption{Double scattering process with moving atoms. The atoms move in an
orthogonal plane respect to the incident wave-vector. We have
$\omega_1=\omega-(\bf{k_i}-\bf{k'})\bf{v_1}$,
$\omega_2=\omega-(\bf{k_i}+\bf{k'})\bf{v_2}$ and
$\omega_{out}=\omega_1-(\bf{k_i}+\bf{k'})\bf{v_2}$. $\bf{k_i}$ ($-\bf{k'}$) is
the incident (intermediate) wave-vector.}
    \label{fig1}
\end{center}
\end{figure}

When the scattering medium is made of moving particles, and this is the case in
an atomic cloud, the amplitudes of the direct and reverse scattering sequences
are no more linked by reciprocity and one observes a dynamical breakdown of the
CBS effect \cite{Golub}. This reduction of the CBS enhancement factor depends
on the velocity distribution of the particles and we can assert very
generally that the
interference contrast will not be much affected provided the velocity spread
will not be too large. For moving resonant scatterers, we will see that ``not
too large'' imposes a stringent condition which will require laser cooling
techniques to be fulfilled.
\\ As a simplistic starting model, we neglect light polarization
effects (scalar wave approximation) and describe atoms as highly resonant
isotropic scatterers. The atomic resonance is characterized by a transition
frequency $\omega_0$ in the optical domain and a transition width $\Gamma$
(typically $\frac{\omega_0}{\Gamma} \simeq 10^8$). The scatterers are then
fully characterized by their complex electrical polarizability
$\alpha(\omega)$, with:
$$\alpha(\omega)=-\frac{3\pi\Gamma c^3}{\omega_0^3}(\frac{1}{\omega-\omega_0+i\frac{\Gamma}{2}})$$
since the scattering differential cross-section is
$\frac{d\sigma}{d\Omega}=\frac{\sigma}{4\pi}$ where the total cross-section is
$\sigma=\frac{k^4|\alpha|^2}{6\pi}$. The velocity distribution of atoms has too
major consequences: first the atomic response at each scattering has to be
evaluated at a frequency $\omega-\bf{kv}$ which is {\it randomly}
Doppler-shifted. This means that each scattering has a variable strength and
that the wave experiences a random phase-kick. Furthermore, after scattering is
completed, the wave has a frequency $\omega'=\omega-(\bf{k}-\bf{k'})\bf{v}$.
Scattering is {\it inelastic} and propagation in an effective medium (until the
following scattering) occurs with an optical index evaluated at random
frequency. As an overall conclusion, there will be a random strength imbalance
and a random dephasing between direct and reverse scattering sequences, leading
to a reduction of the CBS interference. Let us illustrate this dynamical
breakdown for a peculiar double scattering path (see figure~\ref{fig1}). We
consider an incoming resonant light wave (frequency $\omega=\omega_0$,
wave-vector $\bf{k_i}$) and, to simplify the discussion, we only consider the
effect of dephasing between the paths. The total phase shift at backscattering
($\bf{k_{out}}=-\bf{k_i}$) is
$$\Delta\phi=\Delta\phi_{pro}+\Delta\phi_{sca}$$
Where $\Delta\phi_{pro}$ corresponds to the dephasing accumulated during
propagation in the effective medium and $\Delta\phi_{sca}$ the dephasing
accumulated at scattering events. In a dilute medium ($n|\alpha|\ll 1$, where
$n$ is the spatial density), the expressions of $\Delta\phi_{pro}$ et
$\Delta\phi_{sca}$ are quite straightforward:
\begin{equation}
\label{pro}
\Delta\phi_{pro}=\frac{n_r(\omega_1)\omega_1-n_r(\omega_2)\omega_2}{c}L\\
\end{equation}
and
\begin{equation}
\label{sca} \Delta\phi_{sca}=f(\omega-{\bf k_iv_1})+f(\omega_1-{\bf
k'v_2})-f(\omega-{\bf k_iv_2})-f(\omega_2+{\bf k'v_1})
\end{equation}
with
$$f(\omega)=Arg(\alpha)=\arctan(\frac{\Gamma}{2(\omega_0-\omega)})$$
and $$\omega_1=\omega-(\bf{k_i}-\bf{k'})\bf{v_1}$$
$$\omega_2=\omega-(\bf{k_i}+\bf{k'})\bf{v_2}$$
The distance $L$ between scatterers is of the order of the scattering mean free
path $\ell(\omega_0)=\frac{1}{n\sigma(\omega_0)}$. The optical index
$n_r(\omega)$ is given by
$$n_r(\omega)\simeq1+n\frac{Re(\alpha(\omega))}{2}$$
At first order in velocities, we get:
$$\Delta\phi_{pro}\simeq-\frac{1}{\Gamma}((\bf{k'}-\bf{k_i})\bf{v_1}+(\bf{k'}+\bf{k_i})\bf{v_2})$$
and
$$\Delta\phi_{sca}\simeq\frac{4}{\Gamma}\bf{k_i}(\bf{v_1}-\bf{v_2})$$
Averaging over the Gaussian velocity distribution of the two atoms leads to the
interference contrast:
\begin{equation}
\label{con} C(\langle v^2\rangle)=1+\langle
cos(\Delta\phi)\rangle_{\bf{v_1},\bf{v_2}}\\
=1+exp(-6\frac{k^2\langle v^2 \rangle}{\Gamma^2})
\end{equation}
where $$k\simeq\bf{|k_i|}\simeq\bf{|k'|}$$
Expression \ref{con} shows that, to preserve the double scattering CBS effect,
one needs :
\begin{equation}
\label{critere}
\sqrt{\langle v^2\rangle} \ll \Gamma/k
\end{equation}
For scattering sequences of higher
orders, one can argue that the scattering phaseshift roughly performs a random
walk of step of order of $\frac{k\sqrt{\langle v^2\rangle}}{\Gamma}$, thus
increasing as $\sqrt{N}$ with scattering order $N$. This puts a more stringent
condition on the required velocity spread to preserve the CBS effect at higher
orders. For atoms, $\Gamma /k \simeq 10 m/s$, which means that the atomic
gas has first to be laser-cooled before observing CBS. This is done using a
magneto-optical trap (MOT).

\subsection{Atoms cooling and trapping}

\begin{table}[t]
\begin{center}
\begin{tabular}[t]{|c|c|c|c|c|}
\hline Atom
        & $\lambda (nm)$
    & $\frac{\Gamma}{2\pi} (MHz)$
        & $k\sqrt{\langle v^2\rangle}  (\Gamma^{-1})$
            & $C({\langle v^2\rangle})$\\
\hline Rb
    & 780
        & 6
            & $0.04$
                & 2.0\\

\hline Sr
    & 461
        & 32
            & $0.03$
                & 2.0\\
\hline Sr
    & 689
        & 7.10$^{-3}$
            &1
                &\\
\hline He$^*$
    & 1080
        & 1.6
            & 0.2
                & 1.7\\
\hline

\end {tabular}
\caption {Specific values of the double scattering contrast $C({\langle
v^2\rangle})$ for different atoms and transitions. The cases of Rb, Sr and
He$^*$ correspond to dipole-allowed transitions. For the intercombinaison line
cooling of Sr (at $\lambda=689nm$) the transition is too narrow and dynamical
breakdown occurs.} \label{tab}
\end{center}
\end{table}

In a MOT, due to the combined action of the Zeeman and Doppler effects,
a restoring force spatially traps the atoms at the zero of the magnetic
field gradient and cools them in velocity space (for a review on cooling
see {\it e.g.} W. Phillips \cite{nobel}).
The cooling action is based on a Doppler-induced differential
radiation pressure force between contra-propagating laser beams.
For a two-level system in
the low saturation limit and at low velocities ($kv\ll\Gamma$),
the average force
is a pure friction force if the laser is red-detuned with respect to
the transition ($\omega<\omega_0$) . At steady state, this cooling
mechanism is balanced by the heating mechanism induced by the
random character of photon absorption and
emission processes. The equilibrium state is then characterized by
a Gaussian velocity distribution with {\it rms} value:
\begin{equation}
\label{limdop} \sqrt{\langle v^2\rangle}\simeq\sqrt{\frac{\Gamma}{k}v_r}
\end{equation}
where the recoil velocity $v_r =\frac{\hbar k}{m}$ is the velocity change
during a single absorption or emission event. For the usual dipole-allowed
transitions, the recoil frequency shift is small with respect to the width of
the transition
\begin{equation}
\label{fr} kv_r\ll \Gamma
\end{equation}
Thus equation \ref{critere} is well fulfilled in a MOT and the dynamical
breakdown of the CBS effect should be negligible in most cases (see for example
Table \ref{tab}). Moreover, when $F_g>0$, extra cooling mechanics, the
so-called Sisyphus cooling \cite{limdop}, lower the final {\it rms} velocity
with respect to relation (\ref{limdop}).

For Rb, the MOT is loaded in a cell from a low-pressure thermal vapour and for
Sr from an effusive atomic beam. The total number of trapped atoms is
determined by a balance between trapping and loss mechanisms ({\it e.g.}
atom-atom collisions). It usually ranges from $10^7$ to $10^{10}$ atoms in
standard operating conditions. However for localization experiments, it is
crucial to achieve an optical depth at least larger than unity to reach the
multiple scattering regime. This requirement is usually satisfied in a MOT
using a resonant probe beam. It can be also interesting to achieve a high
spatial density for strong localization experiments where $k\ell \simeq 1$ is
demanded. Unfortunately, in most cases, multiple scattering is the dominant
limiting factor \cite{walter} and spatial densities in a MOT are generally
limited to a few of $10^{10}atoms/cm^3$ implying $k\ell\gg 1$.

\section{Reduction of the CBS cone due to internal structure}\label{internal}

\begin{figure}
\begin{center}
\includegraphics[scale=0.5]{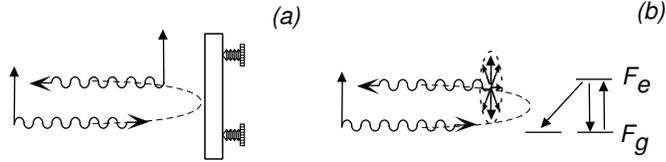}
\caption{For classical spherics scatterer, the backscattered wave conserves his
polarization like in specular reflection off a mirror {\it(a)}. For a quantum
scatterer the polarization is not anymore conserved, due to Raman transition
{\it(b)}.}
\end{center}
    \label{polar}
\end{figure}

In section \ref{vitesse}, we saw how the velocity distribution of
atoms can alter the CBS contrast. Another important suppression
mechanism of the CBS interference effects has been pointed out in reference
\cite{jonck}. It is related to the internal degrees freedom of the scatterers.
In fact, the
optical dipole transition connects two Zeeman-degenerate multiplets (in the
absence of a magnetic field). The groundstate (resp. excited) multiplet
has a total angular momentum $F_g$ (resp. $F_e$) and contains $2F_g+1$
(resp. $2F_e+1$)
magnetic levels. To fully describe scattering, one needs to specify the
initial and final groundstate levels.
Two types of transitions can occur : those leaving the internal state
unaffected (Rayleigh transitions) and those changing
the internal state (Raman transitions).
Thus, when calculating the CBS cone one has to
take properly into account all those allowed mechanisms.
Note that since all levels
in a given multiplet have same energies,
light scattering is always {\it elastic}. However scattering is no more
described by the atomic polarizability alone and one now needs the {\it full
scattering tensor}. In other words, {\it non scalar} features of scattering
will become essential.
\\ It can be shown that the internal structure has
two major consequences. First, because Raman transitions are unavoidable and
since they are accompanied by a change of light polarization at backscattering
(see Figure 2b), the single scattering contribution is no more suppressed in
the $h\|h$ channel. This leads to a rather trivial CBS contrast reduction since
the same would occur with classical {\it non-spherically symmetric} scatterers.
Second, and this point is more subtle, the interfering amplitudes are no more
linked in general by reciprocity. This is because under time-reversal the
angular momentum of atoms has to be flipped. This leads to unbalanced
amplitudes for the direct and reverse scattering sequences and to a decrease of
the CBS contrast as soon as $F_g>0$ \cite{muller01,muller02}. In the $h\|h$
channel, it can be shown that this contrast reduction can be attributed to the
antisymmetric part of the scattering tensor \cite{muller01}. For $F_g=0$
(absence of internal structure in the groundstate), one recovers the previous
classical result. Figure 3 shows the CBS cone for the four usual polarization
channels and for two different transitions in a semi-infinite medium. For the
$F_g=0\rightarrow F_e=1$ transition, cones shape and height are the same than
for a classical dipole-like scatterer. For the $F_g=3\rightarrow F_e=4$
transition, the enhancement factor is systematically lower for all polarization
channels.

\begin{figure}
\begin{center}
\includegraphics[scale=0.5]{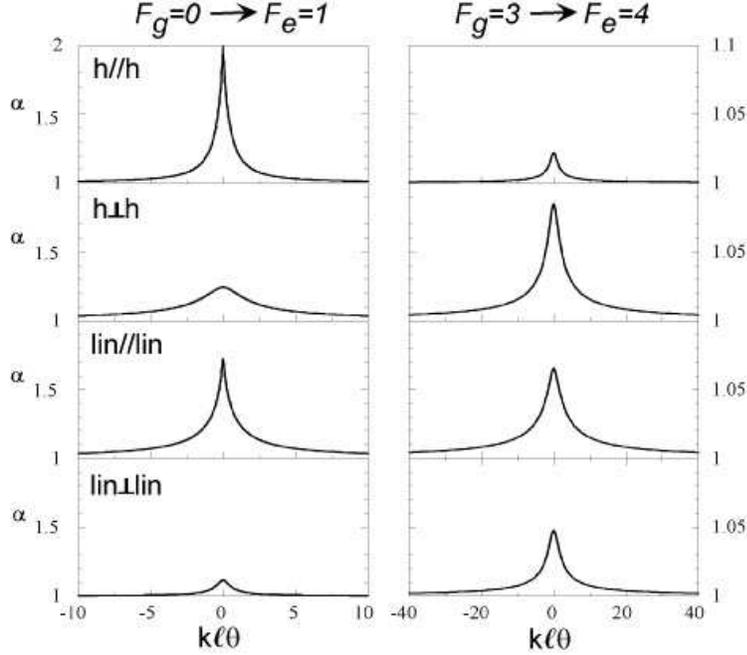}
\caption{CBS cones calculated in a semi-infinite homogeneous scattering medium.
For the $F_g=3\rightarrow F_e=4$ transition, the enhancement factor is
systematical lower for all polarization channels when compared to the
$F_g=0\rightarrow F_e=1$ case where CBS cones are the same than for a classical
dipole-like scatterer.}
\end{center}
    \label{thcbs}
\end{figure}

\section{Experimental results}\label{res}

\begin{figure}[t]
\begin{center}
\includegraphics[scale=0.5]{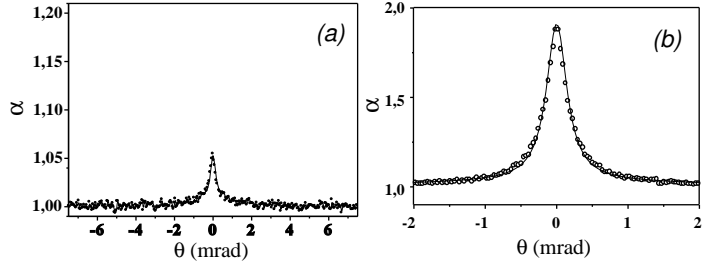}
\caption{Angular dependance of the CBS cone in the helicity preserving $h\|h$
polarization channel. Figure {\it(a)} corresponds to a Rb cloud of 4mm of
diameter and a optical depth of $b\simeq 30$ (the plain line connect two
adjacent experimental points). The CBS cone on Sr is plotted on figure
{\it(b)}. For Sr, the cold cloud has a diameter of 1mm and an optical depth of
$b\simeq 3$. The plain curve in figure {\it (b)} corresponds to a Monte-Carlo
simulation done with the actual experimental parameters.}
\end{center}
    \label{RbSr}
\end{figure}

The detailed experimental procedure for the CBS observation has been published
elsewhere \cite{labeyrie9900}. Briefly, the signal is obtained using a
collimated resonant probe with a beam waist bigger than the size of the cloud
($\simeq 1cm$). To avoid any effects linked to the saturation of the optical
transition (non-linearities, inelastic radiation spectrum) \cite{nonlinear},
the probe intensity is weak (saturation parameter $s\ll 1$). The scattered
light is collected in the backward direction by placing a CCD camera in the
focal plane of an achromatic doublet. The CBS cone plots shown in figure 4 are
obtained in the $h\|h$ polarization. As predicted, the enhancement factor is
strongly reduced for the Rb transition. However a quantitative comparison with
theory calls for some care since the calculation assumes an homogeneous
semi-infinite medium whereas the Rb experiment is made on a finite sample of
Gaussian-distributed atoms. In this type of geometry, it is clear that the
weight of long scattering paths are overestimated with respect to the finite
size sample case. Besides, theoretical calculations show that the major
contribution of the CBS cone comes from low scattering orders. Thus,
considering only the single and double scattering events, the enhancement
factor is predicted to be $\alpha_{th}^{(2)}=1.17$ \cite{jonck}. The
experimental value thus interpolates nicely these two extreme predictions. For
the Sr experiment, the enhancement factor is found to be $\alpha=1.86$ ,
slightly lower the theoretical prediction $\alpha=2$. Several experimental
issues can explain the difference; the finite angular resolution of the
detection apparatus and the imperfect polarization channel isolation
\cite{bidel}. In figure \ref{RbSr}, the experimental Sr CBS cone is compared to
a Monte-Carlo simulation \cite{bidel}. The agreement is clearly excellent.
\section{Conclusion}
To summarize, we have shown that the internal structure and the velocity
distribution of resonant scatterers like atoms have a deep impact on the CBS
effect. The dynamical breakdown induced by the motion of the scatterers can be
made negligible at the expense of using laser-cooled atoms. Unfortunately, the
internal structure irrevocably leads to very small enhancement factors in all
polarization channels as soon as the atomic groundstate is degenerate. This has
been evidenced by experiments on Rubidium. Restoration of a full interference
contrast is obtained with non degenerate atoms like in the Strontium
experiment. This should have interesting potentialities for wave localization
experiments with cold atoms. For example, in the quest for Anderson
localization (which could be obtained only at high density where $kl \approx
1$) where interferences play a crucial role, a $F_{g}=0 \rightarrow F_{e}=1$
transition appears to be a good choice. Is it now possible to increase the
cloud density to reach the Anderson localization threshold? For this purpose,
cooling strontium with the intercombination line in a dipole trap appears to be
a promising technique \cite{katori}.
\section{Acknowledgments}
We gratefully acknowledge D. Delande for stimulating discussions and for
Monte-carlo simulations on the Sr CBS cone. This work is funded by the CNRS and
the PACA region.

\end{document}